\documentclass[sigconf]{acmart}
\renewcommand\footnotetextcopyrightpermission[1]{} 
\usepackage{graphicx}
\usepackage{amsmath}
\usepackage{booktabs}
\usepackage{amsfonts}
\usepackage{multirow}[c]
\usepackage{makecell}
\usepackage{subfig}
\usepackage{color}
\usepackage{bm}
\usepackage{epstopdf}
\usepackage{url}
\usepackage[cal=cm]{mathalfa}
\usepackage{balance}
\usepackage{threeparttable}
\usepackage{wrapfig}
\usepackage{tabularx}
\usepackage{array}
\usepackage{enumitem}
\usepackage{appendix}
\usepackage{tikz}
\usepackage{float} 
\usepackage{adjustbox}
\usepackage[linesnumbered,ruled,vlined]{algorithm2e}

\SetKwProg{Stage}{Stage}{}{} 

\setlist[itemize]{leftmargin=*}

\AtBeginDocument{}

\title{Reward Guided Decoding for Generative Recommendation}

\setcopyright{acmcopyright}
\copyrightyear{2026}
\acmYear{2026}
\acmDOI{XXXXXXX}

\acmConference[Conference acronym 'XX]{Make sure to enter the correct
  conference title from your rights confirmation email}{June 03--05,
  2018}{Woodstock, NY}

\begin{document}

\author{
    Ruochen Yang\textsuperscript{1,2 \dag *}, 
    Yusheng Huang\textsuperscript{3 \dag \S},
    Youfeng Zheng\textsuperscript{4 *},
    Shuang Wen\textsuperscript{3},
    Liangliang Chen\textsuperscript{3}, \\
    Pengbo Xu\textsuperscript{3},
    Xiaoyu Zhang\textsuperscript{3},
    Shijun Wang\textsuperscript{3},
    Shuang Yang\textsuperscript{3 \ddag},
    Zhaojie Liu\textsuperscript{3}, \\
    Lantao Hu\textsuperscript{3},
    Wenwu Ou\textsuperscript{3},
    Jiawei Sheng\textsuperscript{1,2},  
    Tingwen Liu\textsuperscript{1,2 \ddag}
}
\affiliation{
\institution{
    \textsuperscript{1}Institute of Information Engineering, Chinese Academy of Sciences, Beijing, China \\
    \textsuperscript{2}School of Cyber Security, University of Chinese Academy of Sciences, Beijing, China \\ \textsuperscript{3}Kuaishou Technology, Beijing, China
    \textsuperscript{4}Peking University, Beijing, China}
\country{
    \{yangruochen, shengjiawei, liutingwen\}@iie.ac.cn, \{huangyusheng, zhengyoufeng, wenshuang, chenliangliang03, xupengbo03, zhangxiaoyu, wangshijun03, yangshuang08, zhaotianxing, hulantao, luocheng10\}@kuaishou.com
}
}
\thanks{
    * Work done during an internship at Kuaishou Technology. \\
    \dag \ Equal Contribution. \quad
    \ddag \ Corresponding Authors. \quad
    \S \ Project Lead.
}
\renewcommand{\shortauthors}{Ruochen Yang, Yusheng Huang, et al.}

\begin{abstract}
Generative recommendation formulates recommendation task into an SID sequence autoregressive generation paradigm, but the decoding process is often dominated by generation likelihood. 
This may conflict with real-world business objectives, where high-value candidates can receive low generation probability and be pruned early during beam search.
Existing reranking or training-time alignment methods either intervene too late or require costly model retraining when business preferences change.
To this end, we propose \textbf{R}eward \textbf{G}uided \textbf{D}ecoding, named \textbf{RGD}, a controllable decoding framework for industrial value-oriented generative recommendation. 
We formulate value-guided decoding as a KL-regularized reward maximization problem, deriving a closed-form reward guided decoding distribution that principledly combines generation probability with reward signals. 
RGD treats the base generator as a reference policy and introduces a reward model as a test-time controller, injecting reward into each decoding step to reshape the search trajectory without retraining the generator.
Extensive offline and online experiments demonstrate the effectiveness of our approach for aligning personalization and business value.
RGD has been deployed on the Kuaishou platform, bringing consistent improvements in real-world recommendation scenarios.
\end{abstract}



\maketitle

\section{Introduction}

Generative recommendation (GR)~\cite{tiger, onerec} has recently emerged as a highly promising new paradigm in sequential recommendation. Unlike traditional discriminative methods~\cite{sasrec, bert4rec} that retrieve and rank items from a large fixed candidate set by assigning scores, generative recommendation systems draw heavily on the tokenization and generation paradigms of large language models (LLMs)~\cite{grpo}. 
They first represent items into discrete semantic IDs and then generate the next item autoregressively. 
This modeling framework enables end-to-end learning from users’ historical interactions and direct generation of recommendation results, thereby providing users with a fast and personalized experience and demonstrating broad potential in industrial scenarios~\cite{onelive, cobra}.

Despite these advances, most existing generative recommendation methods remain likelihood-driven at their core.
The generator is trained by maximum likelihood estimation over historical interaction sequences, while beam search expands candidates according to local generation probabilities. 
However, in real-world recommendation systems, the item that is most likely under historical behavior patterns is not necessarily the one with the highest business value. 
Industrial platforms typically care about multiple feedback signals, such as clicks, gifts and conversions.
Modeling historical behavior captures users’ natural tendencies under correlation, whereas business systems require the incremental value induced by intervention. 
This discrepancy leads to an inherent mismatch between generation probability and downstream business utility. 
Items with high generation probabilities may reflect user inertia or popularity bias, but this does not imply that low-probability alternatives cannot yield higher user engagement or commercial returns.
This mismatch is further amplified in beam search, where high-reward but low-probability branches are pruned at early search stages, thereby inherently limiting the retrieval space~\cite{promise}.

A natural solution is to introduce reward guidance into the decoding process of generative recommendation. 
However, this is non-trivial for three reasons. 
First, generation probability and reward signals live in different spaces. 
The generator outputs normalized token probabilities, while a reward model may predict click-through rate, gift probability, or a fused multi-objective score.
Directly adding these heterogeneous scores to generation logits is heuristic and lacks a principled interpretation. 
Second, reward guidance should take effect early enough. 
Post-hoc reranking can only reorder completed generative candidates, and therefore cannot recover valuable candidates that have already been eliminated during search. 
Third, industrial objectives are dynamic and various. 
Different scenarios may prefer different trade-offs among click, watch time, gift, and other business metrics. 
Retraining the generator for each preference change is costly and impractical.

Existing approaches only partially address these challenges. 
Conventional generative recommendation methods focus on improving semantic tokenization, sequence modeling, and constrained generation, but still decode mainly according to likelihood~\cite{tiger, nezha, rpg}. 
Reranking-enhanced methods introduce additional rankers or process reward models to filter or reorder generated candidates~\cite{onepiece, rankgr, oneranker, promise}, yet their reward signals are often used as post-hoc validators or pruning heuristics rather than as a principled business-value controller throughout decoding. 
Training-time preference optimization methods utilize RL-based such as DPO or GRPO to align the model with reward signals by updating model parameters~\cite{onerec, onelive}. 
However, such methods bake preferences into the generator, slowing objective switching and reducing flexibility.
Considering reward guided generation have shown that external guidance signals can steer pretrained generators at inference time~\cite{fudge, gedi, prm}. 
Nevertheless, how to formulate and instantiate such principle in generative recommendation with real business feedback remains underexplored.

In this paper, we propose \textbf{R}eward \textbf{G}uided \textbf{D}ecoding, named \textbf{RGD}, a controllable decoding framework for business value-oriented generative recommendation. 
RGD treats the base generator as a reference policy and introduces an additional reward model as a test-time controller. 
At each SID decoding step, the reward model estimates the value of candidate codes under the guidance objectives.
We formulate this process as a KL-regularized reward maximization problem, which naturally yields a closed-form reward guided decoding distribution. 
This provides a principled criterion for combining generation likelihood and reward signals, 
avoiding heuristic score interpolation.
During inference, RGD injects reward into beam search at the SID level, allowing reward to influence the search trajectory before valuable candidates are pruned. 
Moreover, since the reward objective is introduced as a decoding-time controller, RGD supports flexible objective switching and multi-objective preference adjustment without retraining the base generator. 
This makes RGD particularly suitable for industrial scenarios, where the base generator may already be deployed and business objectives need to be adjusted frequently under real-time serving constraints.

We summarize our main contributions as follows:
\begin{itemize}
    \item \textbf{Theoretical Foundation:} We formalize value guidance in generative recommendation as a reward maximization problem. This formulation naturally derives a closed-form reward guided decoding distribution, providing a principled criterion for combining business signals with generative probabilities.
    \item \textbf{Methodological Proposal:} We propose RGD, a controllable reward-guided decoding framework. RGD treats the base generator as a reference policy and introduces an additional reward model as a test-time controller, enabling reward injection at each SID decoding step. It supports dynamic adjustment of multi-objective rewards and weights during serving, allowing flexible alignment and switching across different business preferences.
    \item \textbf{Empirical Validation:} We conduct extensive experiments on both public and industrial datasets, demonstrating significant improvements of RGD over generative recommendation baselines. We further deploy RGD on the Kuaishou platform and verify its business gains under real-world online traffic.
\end{itemize}

\section{Related Work}

\subsection{Generative Recommendation}

Generative recommendation (GR) formulates recommendation as a sequence generation task based on user history. TIGER~\cite{tiger} defines this paradigm with an item tokenization for generating semantic IDs (SIDs) and a T5-based encoder-decoder architecture. 
Following this direction, a series of research works have focused on these two main components.
For item tokenization, Letter~\cite{letter} and MMQ~\cite{mmq} align content semantics with behavioral signals; ETEGRec~\cite{etegrec} and BLOGER~\cite{bloger} explore end-to-end recommendation objective-guided tokenizers; RPG~\cite{rpg} attempts generation of unordered long semantic IDs.
Regarding model architecture, COBRA~\cite{cobra} unifies generation for sparse and dense objectives; EAGER~\cite{eager} and GEMS~\cite{gems} adapt multiple decoders for different tasks; OneLive~\cite{onelive} and Nezha~\cite{nezha} reduce inference latency by introducing sequential multi-token prediction or nimble drafting.

However, most generative recommendation methods are optimized through maximum likelihood-based next-token prediction on historical interactions, which merely fits the exposure item distribution of existing online systems. 
To enable real-world deployment and overcome the cloning of past decisions, the OneRec series~\cite{onerec, oneloc, onemall, onelive} introduced reward models that leverage policy optimization to perceive user preferences. 
They can fundamentally shift the model's generation distribution toward high-reward items, but this black-box policy leads to long-term training and slow transitions for value regulation and business alignment. 
Recent methods have attempted to unlock control over decoding strategies through process rewards. 
V-Star~\cite{v-star} focuses the search space on high-potential prefixes, while PROMISE~\cite{promise} dynamically prunes to prevent semantic drift. 
However, they primarily address the problem of insufficient exploration in beam search, without effectively providing a principled reward-shaped decoding distribution.

\subsection{Reward Guided Generation}

Reward guided generation has been extensively studied in the field of LLMs, with the core objective of guiding pretrained models to produce outputs that better align with human preferences or task-specific rewards. 
One common line of work is training-time reward alignment. RLHF~\cite{rlhf} optimizes the policy by learning a reward model while constraining it from deviating substantially from a reference model. 
Methods such as DPO~\cite{dpo} and GRPO~\cite{grpo} further simplify preference alignment. 
These methods are effective in adjusting the model distribution, but they require additional training to embed preferences into model parameters.
Another line of work focuses on inference-time controlled generation, which keeps the base generator unchanged and introduces external guidance during decoding. 
FUDGE~\cite{fudge} and GeDi~\cite{gedi} use discriminators to guide token generation toward desired attributes. 
Controlled decoding~\cite{controlled_decoding} learns a prefix value function that estimates the expected final reward of partial generations and uses it to reweight decoding probabilities.
PRM~\cite{prm} instead evaluates intermediate reasoning steps with process rewards and prunes low-quality reasoning trajectories during inference. 
These methods offer greater flexibility, as the guidance objective or strength can be adjusted at test time without updating the base generator.
Therefore, incorporating real-world online rewards into the inference decision process of generative recommender systems is highly desirable for value-oriented optimization in industrial applications.

\section{Reward Guided Decoding Formulation}

\subsection{Preliminary}

We first revisit the decoding distribution of a standard generative recommendation. 
Each item has been tokenized in advance into semantic ID composed of
$d$-layer discrete code $[s_1, s_2, \dots, s_d]$.
Given a user context $x$ and a partially generated SID prefix $s_{<l}$, the generator autoregressively predicts the next semantic code $s_l$ over a codebook of size $V$.
Let $\ell_j$ denote the generator logit for candidate code $j$. The corresponding next-token distribution is:
\begin{equation}
p_{\theta}(j \mid s_{<l}, x)
= \frac{\exp(\ell_j)}{\sum_{j'=1}^{V}\exp(\ell_{j'})}.
\end{equation}
For simplicity, we write this distribution as $P(j)$. 
During inference, beam search is employed to make greedy selections of the best top-$K$ combinations step-by-step.
Conventional beam search expands candidates according to $\log P(j)$ and therefore mainly follows the likelihood learned from historical interaction sequences. 

\subsection{Optimal Reward Guided Distribution}

\subsubsection{\textbf{Problem Formulation.}}
In real-world online recommendation, the candidate with the highest generation likelihood is not always the most valuable one. 
Model likelihood mainly captures the probability of reproducing historical interaction patterns, while business value depends on the user feedback after exposure, such as click, long-view, follow, and gift. 
Consequently, decoding may overly select the paths that lead to frequent or easily predictable items but miss candidates with higher downstream value. 
To alleviate this misalignment, we introduce a value function $R(j)$ to explicitly evaluate business feedback and align the decoding process with value-oriented objectives.

Our goal is to construct a new decoding distribution $Q(j)$ that increases expected reward while staying close to the base generator. 
This closeness constraint is important because the generator already captures user preference and item semantics from large-scale behavior data. 
We formulate reward-guided decoding as the following KL-regularized reward maximization problem:
\begin{equation}
Q^* = \arg\max_{Q}
\left\{
\mathbb{E}_{j\sim Q}[R(j)]
- \beta D_{\mathrm{KL}}(Q\|P)
\right\}.
\label{eq:rgd_objective}
\end{equation}
The first term in the formula encourages the selection of codes with higher rewards, while the latter term penalizes the deviation of the new distribution from the existing model.
$\beta>0$ controls the strength of the KL regularization, and its formula is:
\begin{equation}
D_{\mathrm{KL}}(Q\|P)=\sum_j Q(j)\log\frac{Q(j)}{P(j)}.
\end{equation}
This objective treats the base generator as a reference policy and searches for a reward-improved decoding policy. 
A larger $\beta$ forces $Q$ to stay closer to $P$ through more severe punishment, while a smaller $\beta$ allows reward to play a stronger role.

\begin{figure}[t!]
    \centering
    \includegraphics[width=0.95\linewidth]{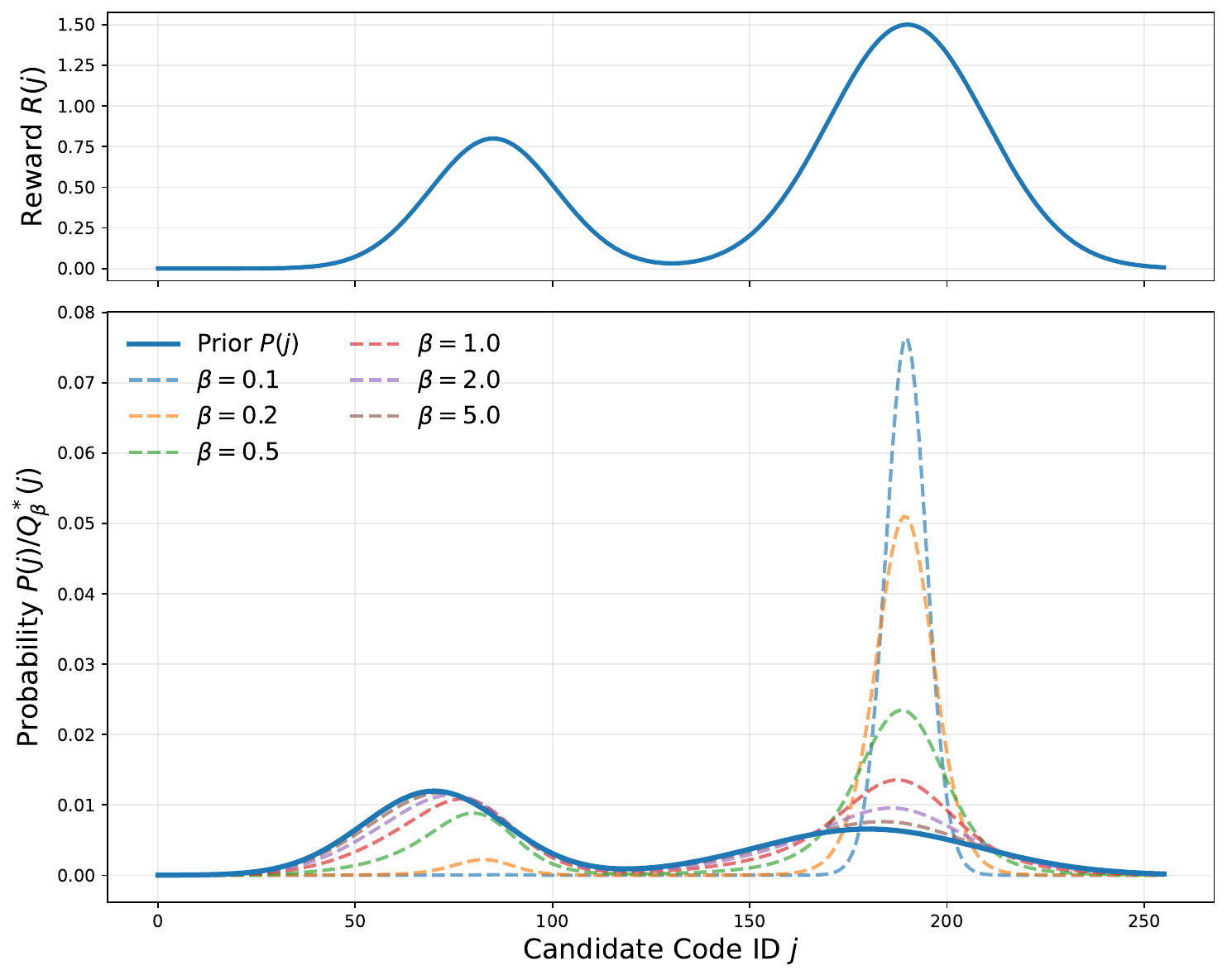}
    \caption{Effect of $\beta$ on the Boltzmann reweighted distribution. Smaller $\beta$ sharpens the distribution toward high-reward candidates, while larger $\beta$ recovers the prior $P$.}
    \label{fig:boltzmann_distributio}
\end{figure}

\begin{figure*}[ht!]
    \centering
    \includegraphics[width=\textwidth]{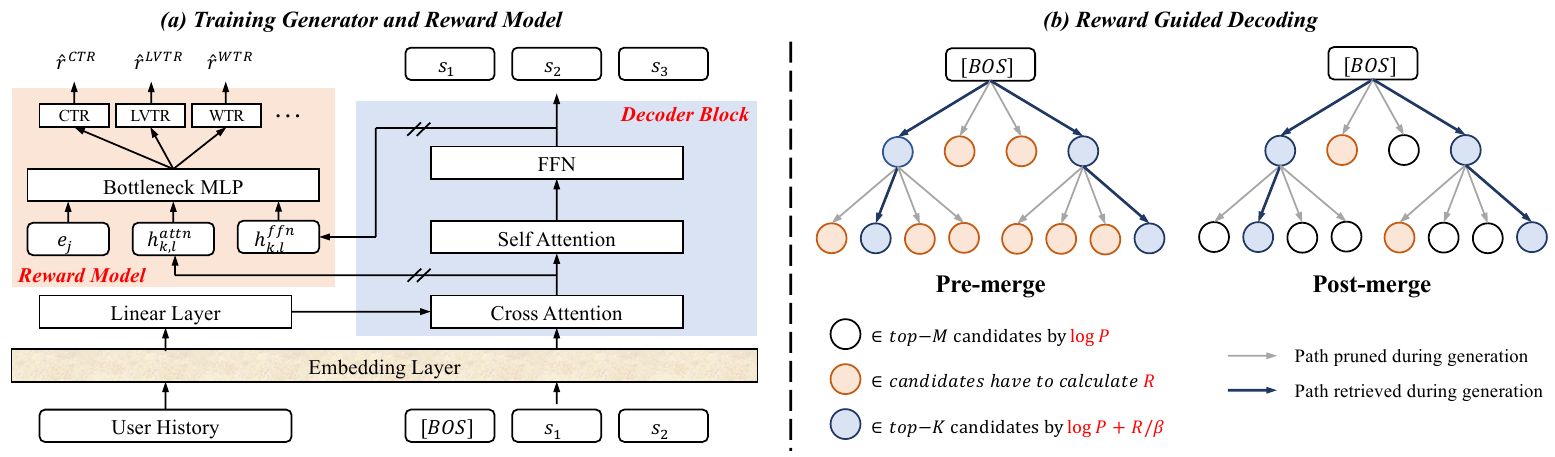}
    \caption{Overall architecture of RGD. (a) The reward model shares the frozen decoder with the base generator, simultaneously calculate the likelihood probability and the reward score. 
    (b) In reward guided decoding, the pre-merge and post-merge inference modes differ in both the number and the location of required reward calculations.
    }
    \label{fig:framework}
\end{figure*}

The objective in Eq.~\ref{eq:rgd_objective} can be viewed as the Lagrangian relaxation of a KL-constrained reward maximization problem. 
Specifically, we could first define the equivalent constraint form of the desired reward-guided distribution as:
\begin{equation}
Q^* = \max_{Q} \mathbb{E}_{j\sim Q}[R(j)]
\quad
\mathrm{s.t.}\quad
D_{\mathrm{KL}}(Q\|P) \leq \epsilon,
\label{eq:kl_constrained_objective}
\end{equation}
which means that we seek a distribution with higher expected reward while restricting its deviation from the base generator. 
The equivalent proof could be found in the Appendix \ref{app:equivalent}.

\subsubsection{\textbf{Closed-Form Solution.}}
We then solve this penalized objective in closed form. Introducing a Lagrange multiplier $\lambda$ for the normalization constraint $\sum_j Q(j)=1$, the Lagrangian is:
\begin{equation}
\mathcal{L}(Q,\lambda)
= \sum_j Q(j)R(j)
- \beta \sum_j Q(j)\log\frac{Q(j)}{P(j)}
- \lambda\left(\sum_j Q(j)-1\right).
\end{equation}
Taking the derivative with respect to $Q(j)$ and setting it to zero gives the first-order optimality condition:
\begin{equation}
\frac{\partial \mathcal{L}}{\partial Q(j)}
= R(j) - \beta\left(\log\frac{Q(j)}{P(j)}+1\right)-\lambda = 0.
\end{equation}
For the right side of the above equation, perform the rearrangement and we obtain:
\begin{equation}
\label{eq:log_q}
\begin{aligned}
    \log Q(j) &= \log P(j) + \frac{R(j)}{\beta} - \frac{\lambda+\beta}{\beta} \\
    &= \log P(j) + \frac{R(j)}{\beta} - \log Z,
\end{aligned}
\end{equation}
where $\log Z=(\lambda+\beta)/\beta$ since the last term is independent of $j$ and thus acts as a normalization constant.
Exponentiating both sides yields the closed-form reward-guided decoding distribution:
\begin{equation}
Q^*(j)=\frac{P(j)\exp(R(j)/\beta)}{Z},
\quad
Z=\sum_{j'}P(j')\exp(R(j')/\beta).
\label{eq:closed_form_q}
\end{equation}

The solution in Eq.~\ref{eq:closed_form_q} has a Boltzmann form:
\begin{equation}
Q^*(j) \propto P(j)\exp(R(j)/\beta),
\end{equation}
which means that the reward-guided distribution is obtained by reweighting the base generator distribution with an exponentially scaled reward term. 
The temperature parameter $\beta$ controls the sharpness of this reweighting, as seen in the Figure \ref{fig:boltzmann_distributio}.

For top-$K$ decoding or beam expansion, we only need to compare the relative order of candidate tokens rather than compute their normalized probabilities. 
Considering that conventional beam search ranks candidates using $\log P(j)$, the relative ranking under the reward-guided distribution is therefore based on $\log Q^*(j)$, \textit{i.e.}, in the form of Eq.\ref{eq:log_q}.
Since the normalization constant $Z$ is shared by all candidates, it does not affect ranking. 
Therefore, selecting the top candidates under $Q^*$ is equivalent to ranking them by:
\begin{equation}
p^*(j)=\log P(j)+\frac{R(j)}{\beta}.
\label{eq:rgd_score}
\end{equation}
This yields a practical decoding criterion for reward-guided generation, replacing heuristic score interpolation with a principled ranking objective derived from the optimal KL-regularized reward-shaped distribution.
The optimality guarantee could be found in the Appendix \ref{app:guarantee}.

\subsection{Connection to Policy Shaping}

Notably, the optimal reward guided distribution is structurally aligned with KL-regularized policy optimization in RLHF~\cite{rlhf}, particularly the PPO-style alignment framework~\cite{ppo}.
Such methods optimize a policy by maximizing expected reward while penalizing deviation from a reference policy:
\begin{equation}
\max_{\pi}
\mathbb{E}_{y\sim \pi(\cdot|x)}[R(y,x)]
-
\beta D_{\mathrm{KL}}(\pi(\cdot|x)\|\pi_{\mathrm{ref}}(\cdot|x)).
\end{equation}
In our formulation, the base generative recommender $P$ corresponds to the reference policy, and the reward guided distribution $Q$ corresponds to the improved policy. Thus, reward guided decoding can be seen as a policy improvement step over the base recommender.

Different from PPO-style training, we do not learn a new parameterized policy. 
Instead, we solve the local policy improvement problem directly during decoding. 
At each SID step, the candidate space is a finite codebook, so the optimal improved distribution has the closed-form Boltzmann form in Eq.~\ref{eq:closed_form_q}. 
This turns policy improvement into a simple decoding-time scoring rule, avoiding online rollout, reward backpropagation, and generator retraining.

This view justifies the practical design of reward-guided decoding. 
The base generator preserves the collaborative and semantic knowledge learned from historical sequences, while the reward term shifts the decoding distribution toward candidates with higher business value, providing a principled knob between likelihood preservation and value maximization.

\section{Methodology}

Building upon the closed-form reward guided distribution derived in Eq.\ref{eq:closed_form_q}, 
we implement it as a practical framework for \textbf{R}eward \textbf{G}uided \textbf{D}ecoding, named \textbf{RGD}.
Figure \ref{fig:framework} illustrates the overall architecture.
Three concrete design choices need to be made: (i) how to parameterize the reward $R(j)$ so that it is computable for every candidate code at every decoding step; (ii) how to train the reward model without interfering with the base generator; and (iii) how to inject the reward into beam search so that guidance takes effect at the right granularity. 

\subsection{Reward Model Design}

The reward model $R_\phi$ is a lightweight scoring head attached to a pretrained generator. Its purpose is to estimate the business value of appending a candidate code $j$ to the current SID prefix, conditioned on the same user context that the generator has already encoded. Two design principles guide our architecture. First, $R_\phi$ should be \emph{representation-aligned} with the generator, so that reward evaluation reuses the semantic space learned during generation rather than reconstructing it. Second, $R_\phi$ should be \emph{chain-structured}, mirroring the autoregressive nature of SID decoding, so that the reward at layer $l$ reflects not only the current candidate but also the entire committed prefix $s_{<l}$.

\subsubsection{\textbf{Generator Feature Extraction.}}
Instead of building an independent tower with its own encoder, we tap directly into the decoder of the base generator.
Since each decoder block contains three residual sub-layers~\cite{onerec-v2}, a cross-attention layer, a self-attention layer and an feed-forward network layer, we extract the hidden outputs from the first and the last at the $l$-th decoding step:
$\mathbf{h}^{\text{attn}}_{k,l}, \mathbf{h}^{\text{ffn}}_{k,l} \in \mathbb{R}^{d_{\text{model}}}$,
where $k$ denotes we select $k$-th decoder block. 
Drawing from empirical observations~\cite{bert4rec, remote} and to accommodate the single-layer MTP decoder block in the base architecture~\cite{onelive}, we set $k=0$. 
These two hidden states are regarded as a comprehensive summary of the context under the current prefix derived from the generator. 
We freeze them to ensure that reward training does not interfere with the generation trajectory.

\subsubsection{\textbf{Chain-Structured Scoring Head.}}
For each candidate code $j$ at layer $l$, we look up the candidate code embedding $\mathbf{e}_j$ from a dedicated table $\mathbf{E}^{\text{r}} \in \mathbb{R}^{V \times d_{model}}$.
The three features are concatenated and projected to a shared context dimension $d_r$:
\begin{equation}
\mathbf{u}_l^{(j)} = \text{ReLU}\big(\mathbf{W}_{\text{in}}\,[\mathbf{e}_j;\,\text{sg}(\mathbf{h}^{\text{attn}}_{k,l});\,\text{sg}(\mathbf{h}^{\text{ffn}}_{k,l})]\big),
\end{equation}
where $\text{sg}(\cdot)$ denotes the stop-gradient operator. 
To let the reward at layer $l$ depend on the committed prefix, we cache the projections $\mathbf{u}_1, \dots, \mathbf{u}_{l-1}$ of the former codes and combine them with the current candidate through a per-layer causal bottleneck MLP~\cite{bottlenneck-mlp}:
\begin{equation}
\mathbf{c}_l^{(j)} = f^{\text{up}}_l\!\left(\text{ReLU}\!\left(f^{\text{down}}_l\!\left([\mathbf{u}_1;\dots;\mathbf{u}_{l-1};\mathbf{u}_l^{(j)}]\right)\right)\right),
\end{equation}

where $f^{\text{down}}_l$ and $f^{\text{up}}_l$ are dimensionality increase and decrease transformations based on a small bottleneck width.
A three-layer MLP head then produces a scalar logit which is squashed to a probability:
\begin{equation}
\tilde r_\phi(j\mid s_{<l},x) = \sigma\!\big(\text{MLP}(\mathbf{c}_l^{(j)})\big) \in (0,1).
\end{equation}

The optimal decoding distribution derived in Eq.~\ref{eq:rgd_score} requires the reward to reside in the same log-space as $\log P(j)$, whereas the scoring head natively outputs a bounded probability $\tilde r_\phi\in(0,1)$. 
We therefore apply a log-odds ratio transform that inverts the final sigmoid  $\tilde r_\phi$ back to an unbounded real-valued signal, so that the fusion becomes dimensionally consistent:
\begin{equation}
R(j\mid s_{<l},x) = \log\frac{\tilde r_\phi(j\mid s_{<l},x)}{1-\tilde r_\phi(j\mid s_{<l},x)}.
\label{eq:logodds}
\end{equation}
Equivalently, $R(j)$ can be read as a log-likelihood ratio distinguished by the reward head between positive and negative feedback,
giving the fused score $p^*(j)$ a Bayesian posterior-update interpretation on top of the generative prior.

\subsection{Reward Model Training}

\subsubsection{\textbf{Training Objective.}}
The reward model is trained jointly with the generator on the same batches, but with strictly separated gradient flow. The total loss is:
\begin{equation}
\mathcal{L}_{\text{total}} = \mathcal{L}_{\text{gen}} + \lambda_r \cdot \mathcal{L}_{\text{r}},
\end{equation}
where $\mathcal{L}_{\text{gen}}$ is the standard next-token cross-entropy of the generator, $\mathcal{L}_{\text{r}}$ is the loss of reward model contributing gradients to components for scoring head,
and $\lambda_r$ balances the two terms. 

\begin{algorithm}[t]
\caption{RGD Beam Search}
\label{alg:rgd_beam}
\SetAlgoLined
\DontPrintSemicolon
\KwIn{user context $x$; beam size $K$; expansion size $M$; mode $\in\{\text{pre},\text{post},\text{hybrid}\}$}
\KwOut{$K$ generated SID sequences}
Initialize beams $\mathcal{B}_0 = \{(\emptyset, 0)\}$; cache $\mathcal{H}=\emptyset$\;
\For{$l = 1, \dots, d$}{
  \ForEach{beam $b\in\mathcal{B}_{l-1}$}{
    Compute $\log P(\cdot\mid b,x)$ and cache $\mathbf{u}_l$\;
    Select the top-$M$ candidates $\mathcal{T}_b$\ by $\log P$ \;
    \If{mode = pre \textnormal{or} (mode = hybrid and $l=1$)}{
      Score every $j\in\mathcal{T}_b$ by Eq.~\ref{eq:rgd_score}\;
    }
  }
  Merge $\bigcup_b \mathcal{T}_b$ into candidate pool $\mathcal{P}$\;
  \If{mode = post \textnormal{or} (mode = hybrid and $l>1$)}{
    Keep an expanded pool $\mathcal{P}'\subseteq\mathcal{P}$ by $\log P$\;
    Score every $j\in\mathcal{P}'$ by Eq.~\ref{eq:rgd_score}\;
    $\mathcal{P}\leftarrow\mathcal{P}'$\;
  }
  Keep top-$K$ candidates as $\mathcal{B}_l$ and reorder caches\;
}
\Return $\mathcal{B}_d$
\end{algorithm}

\subsubsection{\textbf{Single-objective Reward Loss.}}
We first consider single-objective guidance with four business feedback targets: click, long view, follow and gift, denoted as CTR, LVTR, WTR and GTR. 
For each target $m\in\mathcal{M}=\{\mathrm{ctr},\mathrm{lvtr},\mathrm{wtr},\mathrm{gtr}\}$, the reward head predicts a per-level probability by the scoring module $g_m$ above:
\begin{equation}
\hat r^m_l = \sigma\big(g_m(\mathbf{c}_l)\big), \quad l=1,\dots,d,
\end{equation}
where $\mathbf{c}_l$ is the chain-structured representation.
The reward loss for target $m$ is a weighted binary cross-entropy over all SID levels:
\begin{equation}
\mathcal{L}_{\text{r}}^{m}
=
\frac{1}{d}\sum_{l=1}^{d}
\mathrm{BCE}\left(y^m, \hat r^m_l; w^m\right),
\end{equation}
where $y^m$ is the behavior label and $w^m$ is the corresponding sample weight. 
Positive samples are impressions with the target behavior observed, while negatives are valid exposed or watched samples where the target behavior does not occur. This yields separate reward head for guided decoding.

\subsubsection{\textbf{Multi-objective Reward Loss.}}
For multi-objective guidance, we utilize the LTR reward that represents a fused business preference. 
Following the practice of multi-objective ensemble modeling~\cite{pantheon}, the LTR head predicts one scalar reward probability $\hat r^{\mathrm{ltr}}_l$ at each SID level, and is optimized by a weighted aggregation of behavior-level losses:
\begin{equation}
\label{eq:ltr}
\mathcal{L}_{\text{r}}^{\mathrm{ltr}}
=
\sum_{m\in\mathcal{M}} \alpha_m
\frac{1}{d}\sum_{l=1}^{d}
\mathrm{BCE}\left(y^m, \hat r^{\mathrm{ltr}}_l; w^m\right),
\end{equation}
where $\alpha_m$ denotes the business weight of target $m$. Compared with single-objective heads, the LTR reward provides a unified guidance signal that balances multiple objectives and can be directly used as the test-time controller for multi-objective reward guided decoding.

\begin{table*}[ht!]
  \caption{The overall performance comparison of different models on three datasets. The best results are boldfaced and the second-best results are underlined. 
  Results marked with * are obtained from our reproduction experiments\protect\footnotemark, while the remaining results are taken from the corresponding papers.
  All the results of RGD are statistically significant with $p<0.05$ compared to the baselines by paired t-test.}
  \label{tab:overall}
  \begin{tabular}{@{}lcccccccccccc@{}}
    \toprule
    \multirow{2.5}{*}{\textbf{Models}} & \multicolumn{4}{c}{\textbf{Sports and Outdoors}}  & \multicolumn{4}{c}{\textbf{Beauty}} & \multicolumn{4}{c}{\textbf{Toys and Games}}
    \\ \cmidrule(r){2-5} \cmidrule(r){6-9} \cmidrule(r){10-13} & R@5 & N@5 & R@10 & N@10 & R@5 & N@5 & R@10 & N@10 & R@5 & N@5 & R@10 & N@10\\
    
    \midrule
    \midrule
    \multicolumn{13}{c}{\textit{Tranditional}} \\
    \midrule
    GRU4Rec~\cite{gru4rec} & 0.0129 & 0.0086 & 0.0204 & 0.0110 & 0.0164 & 0.0099 & 0.0283 & 0.0137 & 0.0097 & 0.0059 & 0.0176 & 0.0084 \\
    BERT4Rec~\cite{bert4rec} & 0.0115 & 0.0075 & 0.0191 & 0.0099 & 0.0203 & 0.0124 & 0.0347 & 0.0170 & 0.0116 & 0.0071 & 0.0203 & 0.0099 \\
    SASRec~\cite{sasrec} & 0.0233 & 0.0154 & 0.0350 & 0.0192 & 0.0387 & 0.0249 & 0.0605 & 0.0318 & 0.0463 & 0.0306 & 0.0675 & 0.0374 \\
    S$^3$-Rec~\cite{s3-rec} & 0.0251 & 0.0161 & 0.0385 & 0.0204 & 0.0387 & 0.0244 & 0.0647 & 0.0327 & 0.0443 & 0.0294 & 0.0700 & 0.0376 \\
    VQ-Rec~\cite{vq-rec} & 0.0208 & 0.0144 & 0.0300 & 0.0173 & 0.0457 & 0.0317 & 0.0664 & 0.0383 & 0.0497 & 0.0346 & 0.0737 & 0.0423 \\

    \midrule
    \multicolumn{13}{c}{\textit{Generative}} \\
    \midrule
    TIGER~\cite{tiger} & 0.0264 & 0.0181 & 0.0400 & 0.0225 & 0.0454 & 0.0321 & 0.0648 & 0.0384 & \underline{0.0521} & \underline{0.0371} & 0.0712 & \underline{0.0432} \\
    HSTU~\cite{hstu} & 0.0258 & 0.0165 & 0.0414 & 0.0215 & 0.0469 & 0.0314 & 0.0704 & 0.0389 & 0.0433 & 0.0281 & 0.0669 & 0.0357 \\
    RPG*~\cite{rpg} & \underline{0.0281} & \underline{0.0194} & 0.0428 & \underline{0.0241} & \underline{0.0503} & \underline{0.0354} & 0.0714 & \underline{0.0422} & 0.0518 & 0.0356 & \underline{0.0740} & 0.0427  \\
    PROMISE*~\cite{promise} & 0.0271 & 0.0170 & \underline{0.0444} & 0.0226 & 0.0491 & 0.0317 & \underline{0.0775} & 0.0406 & 0.0502 & 0.0340 & 0.0739 & 0.0411  \\
									
    \midrule
    RGD & \textbf{0.0318} & \textbf{0.0204} & \textbf{0.0495} & \textbf{0.0261} & \textbf{0.0550} & \textbf{0.0356} & \textbf{0.0827} & \textbf{0.0445} & \textbf{0.0531} & \textbf{0.0373} & \textbf{0.0766} & \textbf{0.0439}  \\

    \bottomrule
  \end{tabular}
\end{table*}

\subsection{Reward Guided Inference}

During inference, RGD is built upon standard beam search. At each SID layer, beam search maintains the top-$K$ partial SID sequences as active beams. Each beam is expanded with candidate codes from the current codebook, and its top-$M$ candidates are first selected according to the generation probability. The expansions from all beams therefore form a candidate pool containing at most $K\times M$ partial sequences, from which the global top-$K$ sequences are retained for the next layer. RGD introduces reward guidance into this expansion and pruning process, resulting in a modified score as shown in Eq.~\ref{eq:rgd_score}, so that candidates are selected according to both generation plausibility and estimated business value.
Since evaluating rewards for all expanded candidates introduces additional computation, we design three inference strategies with different effectiveness-efficiency trade-offs.

\footnotetext{The RPG uses OpenAI's \texttt{text-embedding-3-large} as the semantic encoder, which is unfairly compared with other baselines that use \texttt{sentence-t5-base}. PROMISE does not provide the available code, so we conduct a replication referring to the paper.}

\begin{figure}[t!]
    \centering
    \includegraphics[width=0.95\linewidth]{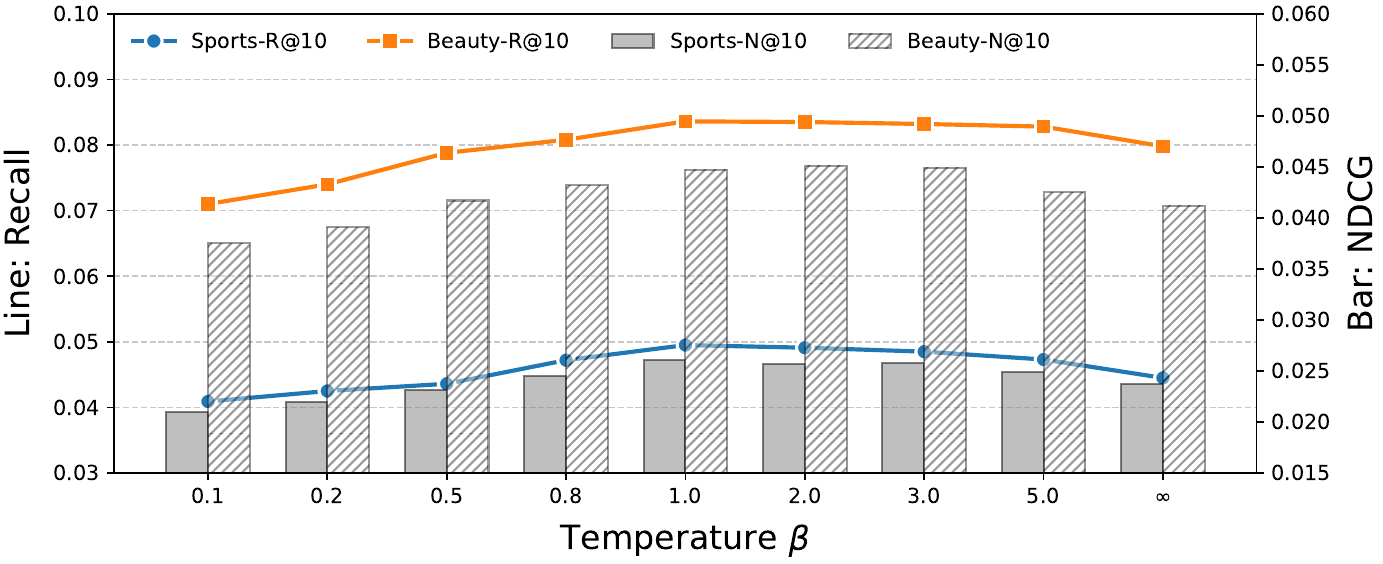}
    \caption{Effect of temperature $\beta$ on Sports and Beauty datasets. $\infty$ is equivalent to no reward guidance.}
    \label{fig:hyper_beta}
\end{figure}

\subsubsection{\textbf{Pre-merge Guidance.}}
In the \emph{pre-merge} mode, each beam evaluates the reward on its own top-$M$ candidates before merging them into the global candidate pool.
Reward therefore influences not only which candidate survives but also which beam is allowed to expand, giving guidance the earliest possible entry point. This is the most expressive but also the most expensive mode, since the reward model is queried $|\mathcal{B}_{l-1}| \cdot M$ times per layer.

\subsubsection{\textbf{Post-merge Guidance.}}
The \emph{post-merge} mode first performs the global merge using only $\log P$, retains an expanded global pool of $M_{\text{expand}} \cdot K$ candidates, and applies the reward only to candidates in this pool. The reward model is called at most $M_{\text{expand}} \cdot K$ times per layer, an order of magnitude fewer than pre-merge. The trade-off is that any candidate whose $\log P$ is not competitive enough to enter the expanded pool cannot be recovered by the reward, even if its business value is high.

\subsubsection{\textbf{Hybrid Guidance.}}
The \emph{hybrid} mode employs pre-merge fusion in the first $l'$ layers and post-merge fusion in the remaining layers.
In our production deployment, we set $l'=1$, since at the first layer there exists only a single active beam. 
This design effectively expands the search space at the first layer which makes critical decisions, while pruning redundant branches in deeper layers, thereby matching the reward-scoring budget we tolerate in production.

Across all three modes, the causal history cache $[\mathbf{u}_1,\dots,\mathbf{u}_{l-1}]$ used by the reward's bottleneck MLP must be reordered together with the generator's key-value cache whenever beams are recombined, so that each surviving beam carries the reward context of its actual ancestor. Both caches are indexed by the same permutation, keeping generation and reward evaluation strictly coupled. The  pseudo code of the algorithm can be seen in Alg.\ref{alg:rgd_beam}.

\section{Experiments}

\begin{table*}[t!]
  \caption{The overall performance comparison of variants driven by different reward methods at the baseline on industrial scenario dataset. The best results are boldfaced.
  \textit{g.} indicates guided by the certain objective.}
  \label{tab:reward}
  \centering
  \begin{tabular}{@{}lcccccc@{}}
    \toprule
    \multirow{2.5}{*}{\textbf{Model}} & \multirow{2.5}{*}{\textbf{HitRate@604}} & \multirow{2.5}{*}{\textbf{MRR@604}} & \multicolumn{4}{c}{\textbf{Reward@604}} \\
    \cmidrule(r){4-7} & & & CTR & LVTR & WTR & GTR \\
    \midrule
    \midrule
    OneLive & 0.82662 & 0.22963 & 0.05442 & 0.05089 & 0.00530 & 0.00155 \\

    \midrule
    DPO & 0.82092 {\scriptsize($-$0.69\%)}
        & 0.22798 {\scriptsize($-$0.72\%)}
        & 0.05495 {\scriptsize($+$0.97\%)}
        & 0.05162 {\scriptsize($+$1.43\%)}
        & 0.00533 {\scriptsize($+$0.57\%)}
        & 0.00156 {\scriptsize($+$0.65\%)} \\
    GRPO & 0.82331 {\scriptsize($-$0.40\%)}
    & 0.22887 {\scriptsize($-$0.33\%)}
    & 0.05548 {\scriptsize($+$1.95\%)}
    & 0.05221 {\scriptsize($+$2.59\%)}
    & 0.00532 {\scriptsize($+$0.38\%)}
    & 0.00157 {\scriptsize($+$1.29\%)} \\

    \midrule
    RGD \textit{g.} CTR
      & 0.82291 {\scriptsize($-$0.37\%)}
      & 0.22955 {\scriptsize($-$0.03\%)}
      & \textbf{0.05618} {\scriptsize(+3.23\%)}
      & 0.05238 {\scriptsize(+2.93\%)}
      & 0.00536 {\scriptsize(+1.13\%)}
      & 0.00157 {\scriptsize(+1.29\%)} \\
    RGD \textit{g.} LVTR
      & 0.82334 {\scriptsize($-$0.33\%)}
      & 0.22894 {\scriptsize($-$0.30\%)}
      & 0.05587 {\scriptsize(+2.58\%)}
      & \textbf{0.05325} {\scriptsize(+4.51\%)}
      & 0.00537 {\scriptsize(+1.31\%)}
      & 0.00156 {\scriptsize(+0.64\%)} \\
    RGD \textit{g.} WTR
      & 0.82336 {\scriptsize($-$0.33\%)}
      & 0.22931 {\scriptsize($-$0.14\%)}
      & 0.05466 {\scriptsize(+0.43\%)}
      & 0.05180 {\scriptsize(+1.71\%)}
      & \textbf{0.00545} {\scriptsize(+2.79\%)}
      & 0.00159 {\scriptsize(+2.56\%)} \\
    RGD \textit{g.} GTR
      & 0.82374 {\scriptsize($-$0.29\%)}
      & 0.22846 {\scriptsize($-$0.51\%)}
      & 0.05408 {\scriptsize($-$0.62\%)}
      & 0.05084 {\scriptsize($-$0.10\%)}
      & 0.00539 {\scriptsize(+1.65\%)}
      & \textbf{0.00164} {\scriptsize(+5.66\%)} \\
    RGD \textit{g.} LTR
      & \textbf{0.83562} {\scriptsize(+1.09\%)}
      & \textbf{0.22970} {\scriptsize(+0.03\%)}
      & 0.05489 {\scriptsize(+0.86\%)}
      & 0.05101 {\scriptsize(+0.24\%)}
      & 0.00534 {\scriptsize(+0.75\%)}
      & 0.00162 {\scriptsize(+4.52\%)} \\
    \bottomrule
  \end{tabular}
\end{table*}

In this section, we conduct extensive experiments offline and online to answer the following questions:
\begin{itemize}
    \item \textbf{RQ1:} How does RGD perform on offline public datasets compared to existing baselines?
    \item \textbf{RQ2:} How does RGD perform after the introduction of rewards in industrial scenarios, and what impact do diverse rewards have?
    \item \textbf{RQ3:} What are the reference settings for RGD in business scenarios, and how do they affect the performance of RGD?
    \item \textbf{RQ4:} How does RGD affect real-world online services?
\end{itemize}

\subsection{Offline Public Experiments (RQ1)}

\subsubsection{\textbf{Experimental Settings.}} 
We conducted experiments on three offline public datasets from the Amazon Reviews datasets~\cite{amazon}: \textbf{Sports and Outdoors (Sports)}, \textbf{Beauty} and \textbf{Toys and Games (Toys)}, to preliminarily verify the performance of our model.
The details of experimental settings could be seen in Appendix \ref{app:offline}.

\subsubsection{\textbf{Overall Performance.}}

We compare RGD with multiple baselines in three offline public datasets, with the results shown in Table \ref{tab:overall}. We can observe that:
\begin{itemize}
    \item RGD achieves the best performance across all three datasets and all evaluation metrics, with the improvements up to 11.49\% in Recall@10 and 8.30\% in NDCG@10. This result supports our argument that likelihood-only SID generation is insufficient for recommendation needs, as beam search is predominantly dominated by local generation probabilities. Compared to PROMISE's direct reranking, RGD reshapes the search space distribution by injecting reward guidance at each decoding step, facilitating the retrieval of candidates more aligned with target preferences.
    \item RGD exhibits the best performance across all three datasets, validating that reward-guided decoding remains effective even on public datasets with only implicit interaction feedback. In other words, reward models trained on positive and negative item signals can still provide more useful preference alignment and value guidance beyond the likelihood probabilities learned by the base generator.
    \item The consistent improvements in both Recall and NDCG indicate that RGD not only retrieves more relevant items but also simultaneously enhances the ranking quality of top-ranked items. This can be attributed to the KL-regularized policy shaping: the base generator provides a reference policy, while the reward model serves as a controller at test time, ensuring effective distribution fine-tuning without significant deviation.
\end{itemize}

\subsubsection{\textbf{Hyperparameter Analysis.}}

We study the effect of the temperature parameter $\beta$, which controls the balance between the base generator and reward guidance, and show the results in Figure \ref{fig:hyper_beta}.
The setting $\beta=\infty$ removes the reward term and degenerates to likelihood-based decoding, leading to worse performance. 
This confirms that the reward model provides useful preference signals beyond the generator probability.
Meanwhile, the performance is best with a moderate $\beta$.
An small $\beta$ causes the reward term to dominate the generation probability, leading to decoding that deviates from reliable semantic structures. 
An overly large $\beta$ weakens the reward term, causing the decoding process to degenerate such that RGD cannot adequately correct the mismatch between generation probability and recommendation preferences. 
Nevertheless, $\beta$ exhibits a similar trend across different datasets, manifesting as a stable and easily interpretable control parameter, which supports the value of RGD as a controllable decoding framework.

\subsection{Industrial Scenario Experiments (RQ2)}

\subsubsection{\textbf{Experimental Settings.}}
Experiments on public datasets lack multi-objective interactions, making them unable to effectively reflect RGD's true performance in real-world industrial environments involving larger data scales, more objective-driven optimization, and more severe exposure bias settings. Therefore, we conduct experiments in industrial scenarios based on Kuaishou live-streaming platform data.
The details of experimental settings could be seen in Appendix \ref{app:industry}.

\subsubsection{Overall Performance.}

The results of OneLive as baseline guided by different reward methods are shown in Table \ref{tab:reward}. We make the following observation:
\begin{itemize}
    \item Both RL and RGD approaches for incorporating reward guidance into generative recommendation models can effectively shift the distribution of generated candidates toward business objectives, thereby improving downstream model recognition and acceptance. 
    This demonstrates that injecting real business feedback into the generation process can effectively optimize the likelihood-based local optimum.
    \item Methods such as RL that optimize preferences through training may impair generation capability to some extent due to distributional drift caused by parameter updates to the base generator.
    And considering its training-time optimization, they are not cost-effective.
    In contrast, RGD achieves a more favorable trade-off between generation quality and reward improvement, as it performs reward shaping solely at the decoding stage with minimal impact on the base generator's performance.
    \item Different guiding rewards under RGD lead to distinct optimization behaviors, fully demonstrating its controllability in industrial scenarios. Single-objective guidance significantly improves individual target rewards, while multi-objective LTR guidance achieves balanced improvements across multiple metrics. 
    Reasonable reward combinations better balance generation quality and commercial value, and RGD provides convenience for business alignment and parameter tuning.
\end{itemize}

\subsection{In-depth Analysis (RQ3)}

\subsubsection{Probability-Reward Mismatch}

\begin{figure}[t!]
    \centering
    \subfloat[Sports]{
        \includegraphics[width=4.3cm]{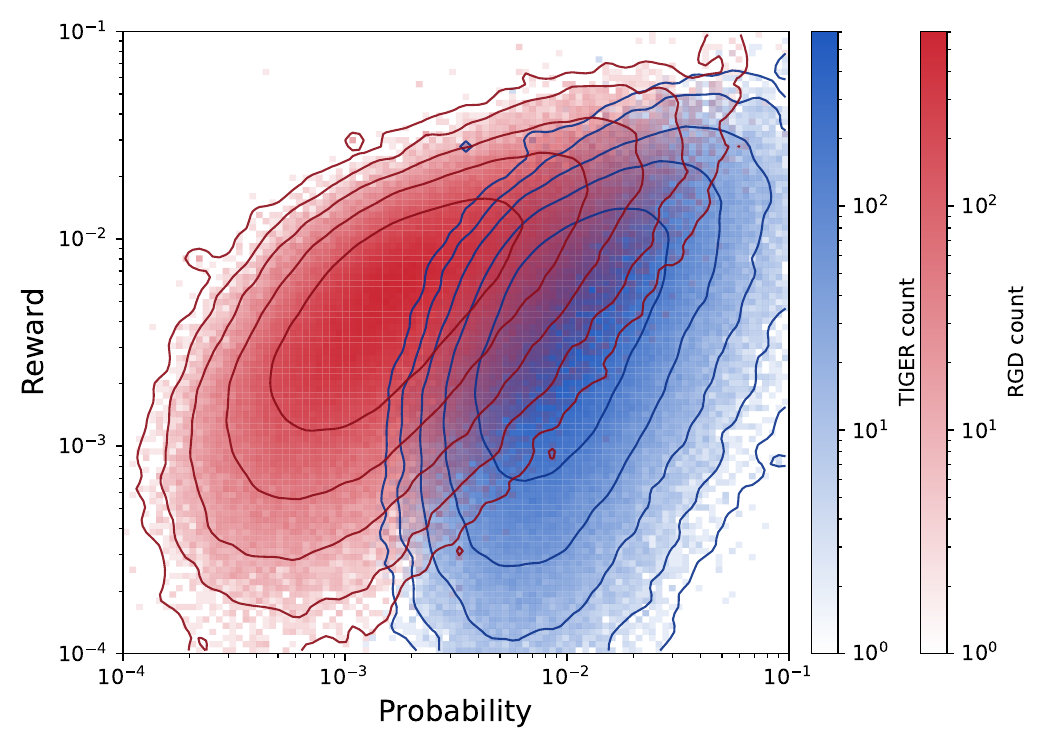}
    }
    \subfloat[Beauty]{
        \includegraphics[width=4.3cm]{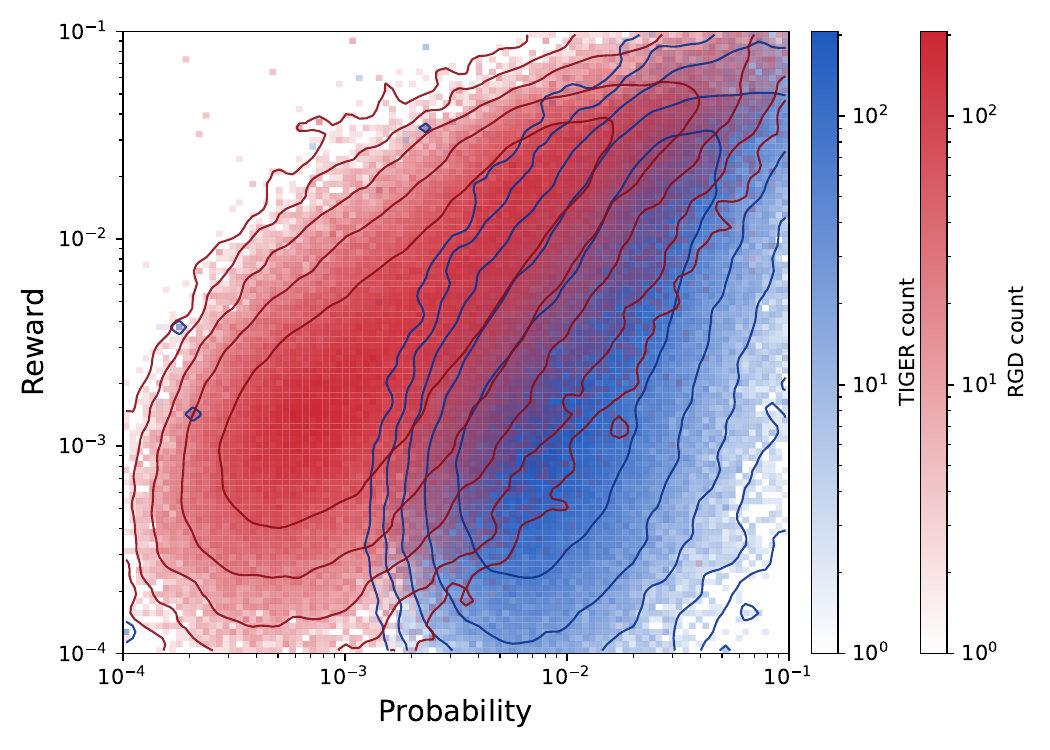}
    }
    \caption{Density distribution of generated candidates in the probability-reward space on Sports and Beauty datasets. The blue contours denote candidates selected by TIGER, while the red contours denote candidates selected by RGD.}
    \vspace{-0.3cm}
    \label{fig:density}
\end{figure}

To understand how RGD changes decoding behavior, we visualize generated candidates in the probability-reward space, where probability is given by the generator and reward is predicted by the reward model. As shown in Figure \ref{fig:density}, TIGER-selected candidates are mainly concentrated in the high-probability region, reflecting its likelihood-dominated beam search strategy.
In contrast, RGD produces a broader distribution and places more density on candidates with relatively lower generation probability but higher reward. 
This indicates that RGD mitigates the probability-reward mismatch by promoting promising candidates that would otherwise be pruned, improving recommendation through reshaping a broader search trajectory and aligning more valuable preferences.

\subsubsection{Inference Mode}

\begin{figure}[t!]
    \centering
    \subfloat[Performance]{
        \includegraphics[width=4.5cm]{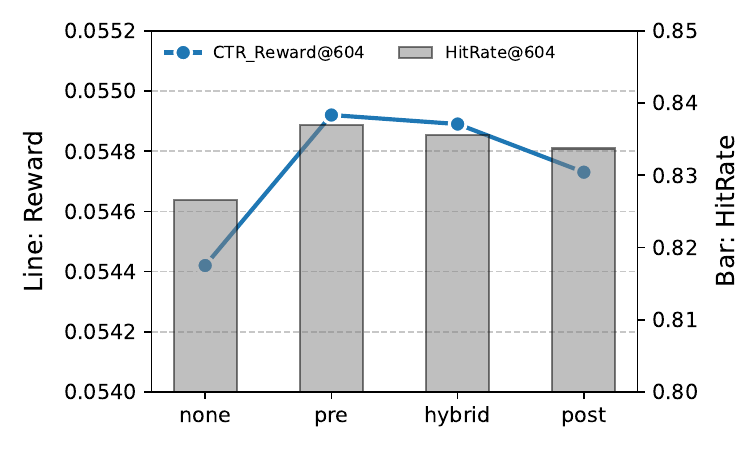}
    }
    \subfloat[Time cost]{
        \includegraphics[width=4.5cm]{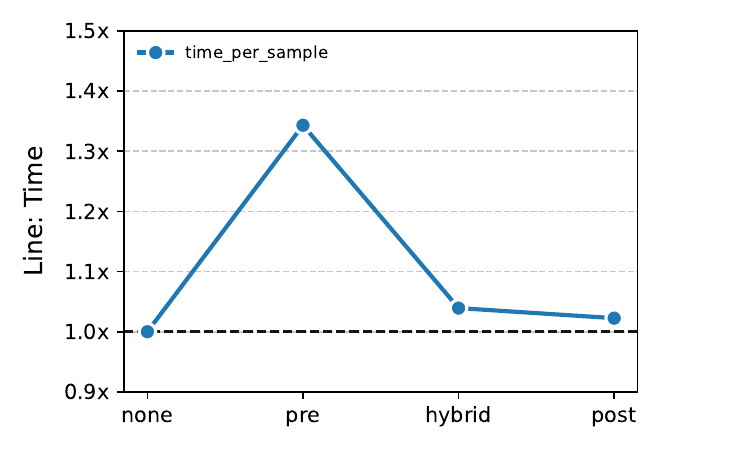}
    }
    \caption{The impact of different inference strategies on (a) performance, and (b) time cost of single sample.
    None indicates baseline without RGD.}
    \label{fig:mode}
\end{figure}

We compared different reward injection strategies during inference under the LTR-guided RGD framework in the industrial scenario\footnotemark, with the results visualized in Figure \ref{fig:mode}. 
The pre-merge strategy achieves the best overall performance, as the larger search space and earlier reward application allow high-reward candidates to influence the search trajectory at an earlier stage.
However, this comes at the cost of higher inference overhead, since rewards must be computed for more candidates. 
In contrast, the hybrid strategy delivers performance comparable to that of pre-merge while incurring latency similar to that of post-merge, and the additional time overhead relative to the no-reward baseline remains acceptable. 
These results indicate that the hybrid strategy offers a more favorable trade-off for production deployment, by conducting fine-grained search over the critical early SIDs followed by beam optimization in the subsequent layers.

\footnotetext{The search space of industrial scenario is substantially larger than that of public datasets, making the distinctions among strategies more pronounced.}

\subsubsection{Reward Model Architecture}

\begin{table}[t]
  \caption{The comparison of results based on different reward model architectures.}
  \label{tab:archi}
  \begin{tabular}{ccc}
    \toprule
    \textbf{Model} & \textbf{CTR\_Reward@604} & \textbf{QPS} \\
    \midrule
    Lite Decoder & - & 1.00x  \\
    Bottleneck MLP & -0.27\% & 1.28x \\
    \bottomrule
  \end{tabular}
\end{table}

We evaluated the performance of different reward model architectures, and the results are presented in the Table \ref{tab:archi}. Lite Decoder indicates the same decoder block as that in the generator. 
It can be observed that even with only a slight acceptable reduction in the CTR reward, the Bottleneck MLP significantly reduces the cost of the inference service. 
Therefore, we choose this architecture as the default reward model.

\subsection{Online A/B Test (RQ4)}

We deploy the CTR-guided version of RGD for online A/B testing on the Kuaishou live-streaming recommendation service. We choose CTR as the guidance objective because click feedback is denser and more stable than other sparse business signals such as follow or gift, making the reward model easier to estimate reliably under real-time serving traffic. In addition, click is an upstream behavior in the live-streaming funnel, and better click guidance can further improve downstream engagement and monetization.

During two-weeks online A/B test, compared with the deployed baseline, RGD improves page-level CTR by \textbf{+0.392\%}, live-streaming watch time by \textbf{+0.689\%} and live-streaming watch counts by \textbf{+0.349\%}. 
These results verify the performance of reward guided decoding for target alignment under real production traffic, as well as the effectiveness of RGD in translating dense user feedback into online business gains.

\section{Conclusion}

In this paper, we presented RGD, a reward guided decoding framework for generative recommendation. We formalized value-guided generation as a KL-regularized reward maximization problem and derived a closed-form reward-shaped distribution that combines generation likelihood with business feedback. 
Based on this formulation, RGD treats the base generator as a reference policy and introduces a reward model as a test-time controller, enabling SID-level reward injection and flexible objective guidance without retraining the generator. 
Experiments on public benchmarks and industrial data demonstrate the  improvements in recommendation accuracy and business rewards, while online A/B tests on Kuaishou confirm its effectiveness in production.

\balance
\bibliographystyle{ACM-Reference-Format}
\bibliography{sample-base-extend.bib}

@inproceedings{bert4rec,
  title={BERT4Rec: Sequential recommendation with bidirectional encoder representations from transformer},
  author={Sun, Fei and Liu, Jun and Wu, Jian and Pei, Changhua and Lin, Xiao and Ou, Wenwu and Jiang, Peng},
  booktitle={Proceedings of the 28th ACM international conference on information and knowledge management},
  pages={1441--1450},
  year={2019}
}

@inproceedings{gru4rec,
  title={When recurrent neural networks meet the neighborhood for session-based recommendation},
  author={Jannach, Dietmar and Ludewig, Malte},
  booktitle={Proceedings of the eleventh ACM conference on recommender systems},
  pages={306--310},
  year={2017}
}

@inproceedings{sasrec,
  title={Self-attentive sequential recommendation},
  author={Kang, Wang-Cheng and McAuley, Julian},
  booktitle={2018 IEEE international conference on data mining (ICDM)},
  pages={197--206},
  year={2018},
  organization={IEEE}
}

@inproceedings{s3-rec,
  title={S3-rec: Self-supervised learning for sequential recommendation with mutual information maximization},
  author={Zhou, Kun and Wang, Hui and Zhao, Wayne Xin and Zhu, Yutao and Wang, Sirui and Zhang, Fuzheng and Wang, Zhongyuan and Wen, Ji-Rong},
  booktitle={Proceedings of the 29th ACM international conference on information \& knowledge management},
  pages={1893--1902},
  year={2020}
}

@inproceedings{vq-rec,
  title={Learning vector-quantized item representation for transferable sequential recommenders},
  author={Hou, Yupeng and He, Zhankui and McAuley, Julian and Zhao, Wayne Xin},
  booktitle={Proceedings of the ACM Web Conference 2023},
  pages={1162--1171},
  year={2023}
}

@article{tiger,
  title={Recommender systems with generative retrieval},
  author={Rajput, Shashank and Mehta, Nikhil and Singh, Anima and Hulikal Keshavan, Raghunandan and Vu, Trung and Heldt, Lukasz and Hong, Lichan and Tay, Yi and Tran, Vinh and Samost, Jonah and others},
  journal={Advances in Neural Information Processing Systems},
  volume={36},
  pages={10299--10315},
  year={2023}
}

@article{hstu,
  title={Actions speak louder than words: Trillion-parameter sequential transducers for generative recommendations},
  author={Zhai, Jiaqi and Liao, Lucy and Liu, Xing and Wang, Yueming and Li, Rui and Cao, Xuan and Gao, Leon and Gong, Zhaojie and Gu, Fangda and He, Michael and others},
  journal={arXiv preprint arXiv:2402.17152},
  year={2024}
}

@inproceedings{rpg,
  title={Generating long semantic ids in parallel for recommendation},
  author={Hou, Yupeng and Li, Jiacheng and Shin, Ashley and Jeon, Jinsung and Santhanam, Abhishek and Shao, Wei and Hassani, Kaveh and Yao, Ning and McAuley, Julian},
  booktitle={Proceedings of the 31st ACM SIGKDD Conference on Knowledge Discovery and Data Mining V. 2},
  pages={956--966},
  year={2025}
}

@article{promise,
  title={PROMISE: Process Reward Models Unlock Test-Time Scaling Laws in Generative Recommendations},
  author={Guo, Chengcheng and Cai, Kuo and Zhou, Yu and Luo, Qiang and Tang, Ruiming and Li, Han and Gai, Kun and Zhou, Guorui},
  journal={arXiv preprint arXiv:2601.04674},
  year={2026}
}

@article{onelive,
  title={OneLive: Dynamically Unified Generative Framework for Live-Streaming Recommendation},
  author={Wang, Shen and Huang, Yusheng and Yang, Ruochen and Wen, Shuang and Xu, Pengbo and Cao, Jiangxia and Liu, Yueyang and Cai, Kuo and Guo, Chengcheng and Wang, Shiyao and others},
  journal={arXiv preprint arXiv:2602.08612},
  year={2026}
}

@inproceedings{letter,
  title={Learnable item tokenization for generative recommendation},
  author={Wang, Wenjie and Bao, Honghui and Lin, Xinyu and Zhang, Jizhi and Li, Yongqi and Feng, Fuli and Ng, See-Kiong and Chua, Tat-Seng},
  booktitle={Proceedings of the 33rd ACM International Conference on Information and Knowledge Management},
  pages={2400--2409},
  year={2024}
}

@inproceedings{mmq,
  title={Mmq: Multimodal mixture-of-quantization tokenization for semantic id generation and user behavioral adaptation},
  author={Xu, Yi and Zhang, Moyu and Li, Chenxuan and Liao, Zhihao and Xing, Haibo and Deng, Hao and Hu, Jinxin and Zhang, Yu and Zeng, Xiaoyi and Zhang, Jing},
  booktitle={Proceedings of the Nineteenth ACM International Conference on Web Search and Data Mining},
  pages={788--797},
  year={2026}
}

@inproceedings{etegrec,
  title={Generative recommender with end-to-end learnable item tokenization},
  author={Liu, Enze and Zheng, Bowen and Ling, Cheng and Hu, Lantao and Li, Han and Zhao, Wayne Xin},
  booktitle={Proceedings of the 48th International ACM SIGIR Conference on Research and Development in Information Retrieval},
  pages={729--739},
  year={2025}
}

@article{bloger,
  title={Bi-Level Optimization for Generative Recommendation: Bridging Tokenization and Generation},
  author={Bai, Yimeng and Liu, Chang and Zhang, Yang and Wang, Dingxian and Yang, Frank and Rabinovich, Andrew and Rong, Wenge and Feng, Fuli},
  journal={arXiv preprint arXiv:2510.21242},
  year={2025}
}

@article{cobra,
  title={Sparse meets dense: Unified generative recommendations with cascaded sparse-dense representations},
  author={Yang, Yuhao and Ji, Zhi and Li, Zhaopeng and Li, Yi and Mo, Zhonglin and Ding, Yue and Chen, Kai and Zhang, Zijian and Li, Jie and LIN, LIU and others},
  journal={Advances in Neural Information Processing Systems},
  volume={38},
  pages={93746--93770},
  year={2026}
}

@article{gems,
  title={Gems: Breaking the long-sequence barrier in generative recommendation with a multi-stream decoder},
  author={Zhou, Yu and Guo, Chengcheng and Cai, Kuo and Liu, Ji and Luo, Qiang and Tang, Ruiming and Li, Han and Gai, Kun and Zhou, Guorui},
  journal={arXiv preprint arXiv:2602.13631},
  year={2026}
}

@inproceedings{eager,
  title={Eager: Two-stream generative recommender with behavior-semantic collaboration},
  author={Wang, Ye and Xun, Jiahao and Hong, Minjie and Zhu, Jieming and Jin, Tao and Lin, Wang and Li, Haoyuan and Li, Linjun and Xia, Yan and Zhao, Zhou and others},
  booktitle={Proceedings of the 30th ACM SIGKDD Conference on Knowledge Discovery and Data Mining},
  pages={3245--3254},
  year={2024}
}

@inproceedings{nezha,
  title={Nezha: A zero-sacrifice and hyperspeed decoding architecture for generative recommendations},
  author={Wang, Yejing and Zhou, Shengyu and Lu, Jinyu and Liu, Ziwei and Liu, Langming and Wang, Maolin and Zhang, Wenlin and Li, Feng and Su, Wenbo and Wang, Pengjie and others},
  booktitle={Proceedings of the ACM Web Conference 2026},
  pages={8073--8082},
  year={2026}
}

@article{onerec,
  title={Onerec: Unifying retrieve and rank with generative recommender and iterative preference alignment},
  author={Deng, Jiaxin and Wang, Shiyao and Cai, Kuo and Ren, Lejian and Hu, Qigen and Ding, Weifeng and Luo, Qiang and Zhou, Guorui},
  journal={arXiv preprint arXiv:2502.18965},
  year={2025}
}

@article{onerec-v2,
  title={Onerec-v2 technical report},
  author={Zhou, Guorui and Hu, Hengrui and Cheng, Hongtao and Wang, Huanjie and Deng, Jiaxin and Zhang, Jinghao and Cai, Kuo and Ren, Lejian and Ren, Lu and Yu, Liao and others},
  journal={arXiv preprint arXiv:2508.20900},
  year={2025}
}

@inproceedings{oneloc,
  title={Oneloc: Geo-aware generative recommender systems for local life service},
  author={Wei, Zhipeng and Cai, Kuo and She, Junda and Chen, Jie and Chen, Minghao and Zeng, Yang and Luo, Qiang and Zeng, Wencong and Tang, Ruiming and Gai, Kun and others},
  booktitle={Proceedings of the Nineteenth ACM International Conference on Web Search and Data Mining},
  pages={735--744},
  year={2026}
}

@article{onemall,
  title={OneMall: One Model, More Scenarios--End-to-End Generative Recommender Family at Kuaishou E-Commerce},
  author={Zhang, Kun and Zhang, Jingming and Cheng, Wei and Cheng, Yansong and Zhang, Jiaqi and Lu, Hao and Zhang, Xu and Gan, Haixiang and Cao, Jiangxia and Wang, Tenglong and others},
  journal={arXiv preprint arXiv:2601.21770},
  year={2026}
}

@article{v-star,
  title={Spend Search Where It Pays: Value-Guided Structured Sampling and Optimization for Generative Recommendation},
  author={Jiang, Jie and Huang, Yangru and Wang, Zeyu and Wang, Changping and Xiong, Yuling and Zhang, Jun and Yu, Huan},
  journal={arXiv preprint arXiv:2602.10699},
  year={2026}
}

@article{rlhf,
  title={Training language models to follow instructions with human feedback},
  author={Ouyang, Long and Wu, Jeffrey and Jiang, Xu and Almeida, Diogo and Wainwright, Carroll and Mishkin, Pamela and Zhang, Chong and Agarwal, Sandhini and Slama, Katarina and Ray, Alex and others},
  journal={Advances in neural information processing systems},
  volume={35},
  pages={27730--27744},
  year={2022}
}

@article{ppo,
  title={Proximal policy optimization algorithms},
  author={Schulman, John and Wolski, Filip and Dhariwal, Prafulla and Radford, Alec and Klimov, Oleg},
  journal={arXiv preprint arXiv:1707.06347},
  year={2017}
}

@inproceedings{amazon,
  title={Justifying recommendations using distantly-labeled reviews and fine-grained aspects},
  author={Ni, Jianmo and Li, Jiacheng and McAuley, Julian},
  booktitle={Proceedings of the 2019 conference on empirical methods in natural language processing and the 9th international joint conference on natural language processing (EMNLP-IJCNLP)},
  pages={188--197},
  year={2019}
}

@inproceedings{fatsmb,
  title={From Agnostic to Specific: Latent Preference Diffusion for Multi-Behavior Sequential Recommendation},
  author={Yang, Ruochen and Li, Xiaodong and Sheng, Jiawei and Cao, Jiangxia and Lin, Xinkui and Wang, Shen and Yang, Shuang and Liu, Zhaojie and Liu, Tingwen},
  booktitle={Proceedings of the 32nd ACM SIGKDD Conference on Knowledge Discovery and Data Mining V. 1},
  pages={1775--1786},
  year={2026}
}

@inproceedings{sentence-t5,
  title={Sentence-t5: Scalable sentence encoders from pre-trained text-to-text models},
  author={Ni, Jianmo and Abrego, Gustavo Hernandez and Constant, Noah and Ma, Ji and Hall, Keith and Cer, Daniel and Yang, Yinfei},
  booktitle={Findings of the association for computational linguistics: ACL 2022},
  pages={1864--1874},
  year={2022}
}

@inproceedings{rq-vae,
  title={Autoregressive image generation using residual quantization},
  author={Lee, Doyup and Kim, Chiheon and Kim, Saehoon and Cho, Minsu and Han, Wook-Shin},
  booktitle={Proceedings of the IEEE/CVF conference on computer vision and pattern recognition},
  pages={11523--11532},
  year={2022}
}

@inproceedings{qarm,
  title={Qarm: Quantitative alignment multi-modal recommendation at kuaishou},
  author={Luo, Xinchen and Cao, Jiangxia and Sun, Tianyu and Yu, Jinkai and Huang, Rui and Yuan, Wei and Lin, Hezheng and Zheng, Yichen and Wang, Shiyao and Hu, Qigen and others},
  booktitle={Proceedings of the 34th ACM International Conference on Information and Knowledge Management},
  pages={5915--5922},
  year={2025}
}

@article{moment,
  title={Moment\&Cross: Next-Generation Real-Time Cross-Domain CTR Prediction for Live-Streaming Recommendation at Kuaishou},
  author={Cao, Jiangxia and Wang, Shen and Li, Yue and Wang, Shenghui and Tang, Jian and Wang, Shiyao and Yang, Shuang and Liu, Zhaojie and Zhou, Guorui},
  journal={arXiv preprint arXiv:2408.05709},
  year={2024}
}

@article{dpo,
  title={Direct preference optimization: Your language model is secretly a reward model},
  author={Rafailov, Rafael and Sharma, Archit and Mitchell, Eric and Manning, Christopher D and Ermon, Stefano and Finn, Chelsea},
  journal={Advances in neural information processing systems},
  volume={36},
  pages={53728--53741},
  year={2023}
}

@article{grpo,
  title={Deepseekmath: Pushing the limits of mathematical reasoning in open language models},
  author={Shao, Zhihong and Wang, Peiyi and Zhu, Qihao and Xu, Runxin and Song, Junxiao and Bi, Xiao and Zhang, Haowei and Zhang, Mingchuan and Li, YK and Wu, Yang and others},
  journal={arXiv preprint arXiv:2402.03300},
  year={2024}
}

@inproceedings{remote,
  title={REMOTE: A Unified Multimodal Relation Extraction Framework with Multilevel Optimal Transport and Mixture-of-Experts},
  author={Lin, Xinkui and Xu, Yongxiu and Tang, Minghao and Zhang, Shilong and Xu, Hongbo and Xu, Hao and Wang, Yubin},
  booktitle={Proceedings of the 33rd ACM International Conference on Multimedia},
  pages={121--130},
  year={2025}
}

@inproceedings{bottlenneck-mlp,
  title={Parameter-efficient transfer learning for NLP},
  author={Houlsby, Neil and Giurgiu, Andrei and Jastrzebski, Stanislaw and Morrone, Bruna and De Laroussilhe, Quentin and Gesmundo, Andrea and Attariyan, Mona and Gelly, Sylvain},
  booktitle={International conference on machine learning},
  pages={2790--2799},
  year={2019},
  organization={PMLR}
}

@inproceedings{pantheon,
  title={Pantheon: Personalized multi-objective ensemble sort via iterative pareto policy optimization},
  author={Cao, Jiangxia and Xu, Pengbo and Cheng, Yin and Guo, Kaiwei and Tang, Jian and Wang, Shijun and Leng, Dewei and Yang, Shuang and Liu, Zhaojie and Niu, Yanan and others},
  booktitle={Proceedings of the 34th ACM International Conference on Information and Knowledge Management},
  pages={5575--5582},
  year={2025}
}

@article{onepiece,
  title={Onepiece: Bringing context engineering and reasoning to industrial cascade ranking system},
  author={Dai, Sunhao and Tang, Jiakai and Wu, Jiahua and Wang, Kun and Zhu, Yuxuan and Chen, Bingjun and Hong, Bangyang and Zhao, Yu and Fu, Cong and Wu, Kangle and others},
  journal={arXiv preprint arXiv:2509.18091},
  year={2025}
}

@article{rankgr,
  title={RankGR: Rank-Enhanced Generative Retrieval with Listwise Direct Preference Optimization in Recommendation},
  author={Fu, Kairui and Wu, Changfa and Yuan, Kun and Cao, Binbin and Huang, Dunxian and Yan, Yuliang and Zheng, Junjun and Zhang, Jianning and Zhou, Silu and Wu, Jian and others},
  journal={arXiv preprint arXiv:2602.08575},
  year={2026}
}

@article{oneranker,
  title={OneRanker: Unified Generation and Ranking with One Model in Industrial Advertising Recommendation},
  author={Sun, Dekai and Liu, Yiming and Zhou, Jiafan and Liu, Xun and Yu, Chenchen and Li, Yi and Zhang, Jun and Yu, Huan and Jiang, Jie},
  journal={arXiv preprint arXiv:2603.02999},
  year={2026}
}

@inproceedings{fudge,
  title={FUDGE: Controlled text generation with future discriminators},
  author={Yang, Kevin and Klein, Dan},
  booktitle={Proceedings of the 2021 Conference of the North American Chapter of the Association for Computational Linguistics: Human Language Technologies},
  pages={3511--3535},
  year={2021}
}

@inproceedings{gedi,
  title={Gedi: Generative discriminator guided sequence generation},
  author={Krause, Ben and Gotmare, Akhilesh Deepak and McCann, Bryan and Keskar, Nitish Shirish and Joty, Shafiq and Socher, Richard and Rajani, Nazneen Fatema},
  booktitle={Findings of the Association for Computational Linguistics: EMNLP 2021},
  pages={4929--4952},
  year={2021}
}

@inproceedings{prm,
  title={Let's verify step by step},
  author={Lightman, Hunter and Kosaraju, Vineet and Burda, Yuri and Edwards, Harrison and Baker, Bowen and Lee, Teddy and Leike, Jan and Schulman, John and Sutskever, Ilya and Cobbe, Karl},
  booktitle={International Conference on Learning Representations},
  volume={2024},
  pages={39578--39601},
  year={2024}
}

@article{controlled_decoding,
  title={Controlled decoding from language models},
  author={Mudgal, Sidharth and Lee, Jong and Ganapathy, Harish and Li, YaGuang and Wang, Tao and Huang, Yanping and Chen, Zhifeng and Cheng, Heng-Tze and Collins, Michael and Strohman, Trevor and others},
  journal={arXiv preprint arXiv:2310.17022},
  year={2023}
}

\appendix

\section{Formula Proof}

\subsection{Equivalent Proof}
\label{app:equivalent}

The KL-regularized objective in the main text can be viewed as the Lagrangian relaxation of the constrained problem in Eq.\ref{eq:kl_constrained_objective}.
Its Lagrangian is:
\begin{equation}
\begin{aligned}
\mathcal{L}(Q,\beta)
& = \mathbb{E}_{j\sim Q}[R(j)] - \beta\left(D_{\mathrm{KL}}(Q\|P)-\epsilon\right) \\
& = \mathbb{E}_{j\sim Q}[R(j)] - \beta D_{\mathrm{KL}}(Q\|P) + \beta\epsilon
,
\quad \beta \geq 0.
\end{aligned}
\end{equation}
For a fixed $\beta$, the last term $\beta\epsilon$ is independent of $Q$, so maximizing $\mathcal{L}(Q,\beta)$ over $Q$ is equivalent to solving:
\begin{equation}
\max_{Q}
\left\{
\mathbb{E}_{j\sim Q}[R(j)]
-
\beta D_{\mathrm{KL}}(Q\|P)
\right\}.
\end{equation}
Thus, the penalized objective is the Lagrangian relaxation of the KL-constrained problem. For each $\beta$, the solution corresponds to a constrained problem with:
\begin{equation}
\epsilon = D_{\mathrm{KL}}(Q^*_{\beta}\|P).
\end{equation}

\subsection{Optimality Guarantee}
\label{app:guarantee}

We further show that the closed-form distribution derived in the main text is optimal under the KL-constrained view. Assume that $P(j)>0$ for all candidates and $R(j)$ is bounded. For a fixed $\beta>0$, the objective is:
\begin{equation}
\mathcal{F}_{\beta}(Q)
=
\mathbb{E}_{j\sim Q}[R(j)]
-
\beta D_{\mathrm{KL}}(Q\|P)
\end{equation}
It is strictly concave over the probability simplex, since the expected reward term is linear in $Q$ and $-D_{\mathrm{KL}}(Q\|P)$ is strictly concave. Therefore, its maximizer is unique. As derived in the main text, this maximizer is:
\begin{equation}
Q^*_{\beta}(j)
=
\frac{P(j)\exp(R(j)/\beta)}{Z_{\beta}}.
\end{equation}

Let $\epsilon_{\beta}=D_{\mathrm{KL}}(Q^*_{\beta}\|P)$. We prove that $Q^*_{\beta}$ is also the unique solution of the constrained reward maximization problem:
\begin{equation}
\max_{Q}\mathbb{E}_{j\sim Q}[R(j)]
\quad
\mathrm{s.t.}
\quad
D_{\mathrm{KL}}(Q\|P)\leq \epsilon_{\beta}.
\end{equation}
Since $Q^*_{\beta}$ maximizes $\mathcal{F}_{\beta}$, for any feasible $Q$ we have:
\begin{equation}
\mathbb{E}_{Q^*_{\beta}}[R]
-
\beta\epsilon_{\beta}
\geq
\mathbb{E}_{Q}[R]
-
\beta D_{\mathrm{KL}}(Q\|P).
\end{equation}
Rearranging gives:
\begin{equation}
\mathbb{E}_{Q^*_{\beta}}[R]
\geq
\mathbb{E}_{Q}[R]
+
\beta\left(\epsilon_{\beta}-D_{\mathrm{KL}}(Q\|P)\right)
\geq
\mathbb{E}_{Q}[R].
\end{equation}
Thus, no distribution within the same KL budget can achieve a higher expected reward. The uniqueness follows from the strict concavity of $\mathcal{F}_{\beta}$.

Equivalently, define $r_{\beta}=\mathbb{E}_{Q^*_{\beta}}[R]$. Then $Q^*_{\beta}$ is also the unique solution of:
\begin{equation}
\min_{Q}D_{\mathrm{KL}}(Q\|P)
\quad
\mathrm{s.t.}
\quad
\mathbb{E}_{j\sim Q}[R(j)]\geq r_{\beta}.
\end{equation}
For any $Q$ satisfying $\mathbb{E}_{Q}[R]\geq r_{\beta}$, the optimality of $Q^*_{\beta}$ for $\mathcal{F}_{\beta}$ implies:
\begin{equation}
r_{\beta}-\beta\epsilon_{\beta}
\geq
\mathbb{E}_{Q}[R]-\beta D_{\mathrm{KL}}(Q\|P)
\geq
r_{\beta}-\beta D_{\mathrm{KL}}(Q\|P),
\end{equation}
which leads to $D_{\mathrm{KL}}(Q\|P)\geq \epsilon_{\beta}$. Hence, among all distributions achieving at least the same expected reward, $Q^*_{\beta}$ is the closest one to the base generator distribution $P$ in KL divergence. This provides the optimality guarantee for RGD: it improves reward as much as possible under a specified deviation budget, or equivalently stays as close as possible to the generator under a specified reward target.

\section{Effect Analysis}
\label{app:beta_entropy}

\begin{figure}[h]
    \centering
    \includegraphics[width=0.9\linewidth]{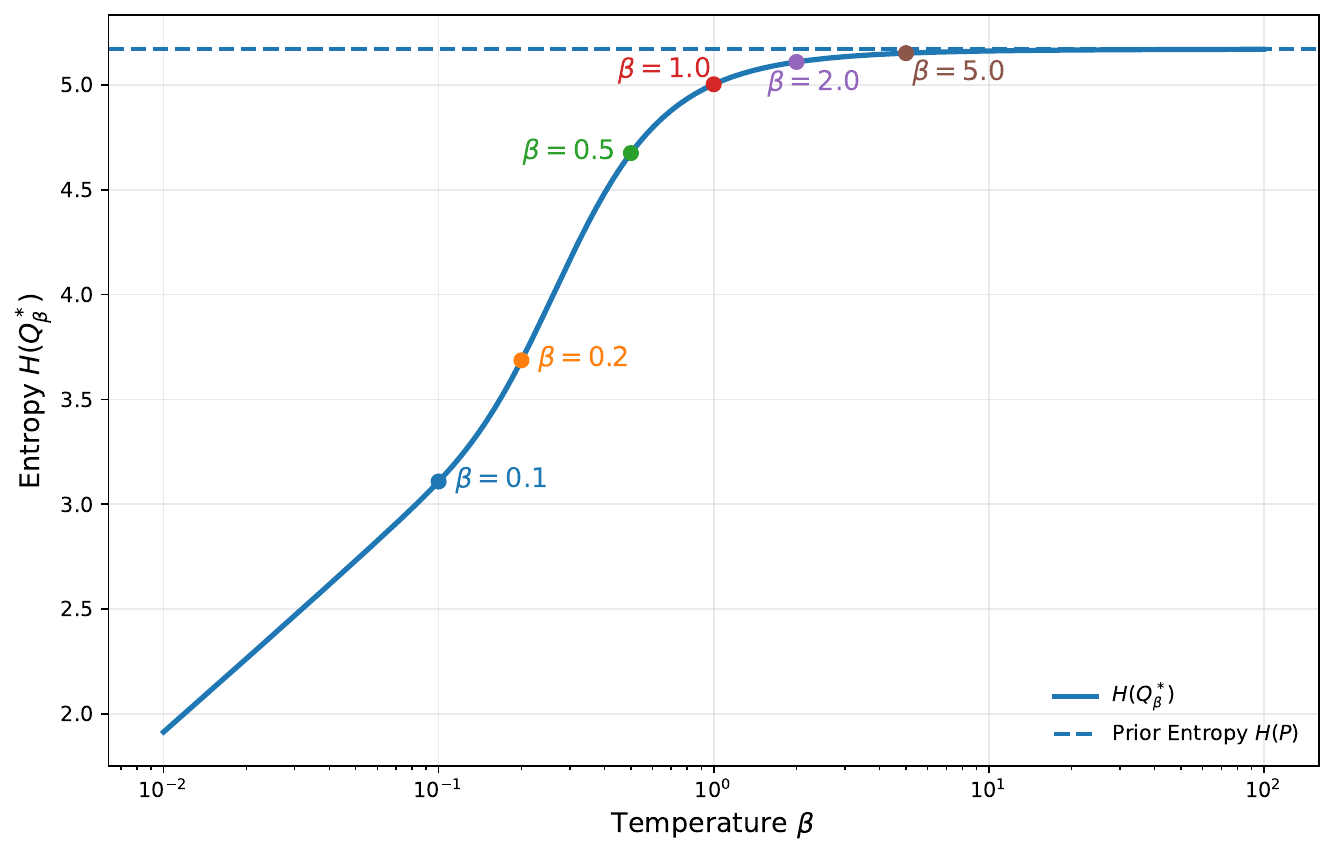}
    \caption{Effect of $\beta$ on the distribution entropy.}
    \label{fig:beta_entropy}
\end{figure}

In addition to the reweighted distribution shown in Figure~\ref{fig:boltzmann_distributio}, we further visualize how the entropy of the reward-shaped distribution changes with the temperature parameter $\beta$. Specifically, we compute the reshaped distribution:
\begin{equation}
Q_\beta^*(j)=\mathrm{softmax}\left(\ell_j+\frac{R(j)}{\beta}\right),
\end{equation}
and plot the entropy $H(Q_\beta^*)$ under different values of $\beta$.

As shown in Figure~\ref{fig:beta_entropy}, the entropy of $Q_\beta^*$ increases monotonically with $\beta$ and gradually approaches the entropy of the prior distribution $P$. This is consistent with the role of $\beta$ in KL-regularized reward shaping. 
Therefore, $\beta$ can be interpreted as a controllable knob that adjusts the sharpness of reward guidance: small $\beta$ leads to stronger and more selective reward-driven decoding, while large $\beta$ recovers likelihood-driven generation.
This also explains the empirical sensitivity of RGD to $\beta$ in the main experiments, as a moderate value of $\beta$ provides a better balance between preserving the base generator distribution and emphasizing high-reward candidates.

\section{Experimental details}

\subsection{Offline Public Experiment Settings}
\label{app:offline}

\textbf{Datasets.}
We evaluate our model on three categories from the Amazon Reviews datasets~\cite{amazon}: \textbf{Sports and Outdoors (Sports)}, \textbf{Beauty} and \textbf{Toys and Games (Toys)}, which are widely used benchmarks for sequential and generative recommendation.
Following standard practice in prior work~\cite{bert4rec, tiger, rpg}, we treat users' historical reviews as interactions and sort them chronologically for input sequences.
We adopt the leave-last-out strategy~\cite{fatsmb, promise} to split the data into training, validation and test sets.
The statistics of the dataset can be seen in Table \ref{tab:data}.

\begin{table}[h]
    \caption{The statistics of the preprocessed datasets, where AvgLen indicates the average length of input sequences}
    \label{tab:data}
    \centering
    \begin{tabular}{cccccc}
        \toprule
        \textbf{Dataset} & \textbf{User} & \textbf{Item} & \textbf{Interaction} & \textbf{Sparsity} & \textbf{AvgLen} \\
        \midrule
        Sports & 35598 & 18357 & 296337 & 99.95\% & 8.32 \\
        Beauty & 22363 & 12101 & 198502 & 99.93\% & 8.88 \\
        Toys & 19412 & 11924 & 167597 & 99.93\% & 8.63 \\
        \bottomrule
    \end{tabular}
\end{table}

\textbf{Baselines.}
We compare our model with the state-of-the-art models from two lines of research topics: 
Traditional Sequential Recommendation Models include: GRU4Rec~\cite{gru4rec}, BERT4Rec~\cite{bert4rec}, SASRec~\cite{sasrec}, S$^3$-Rec~\cite{s3-rec} and VQ-Rec~\cite{vq-rec}.
Generative Recommendation Models include: TIGER~\cite{tiger}, HSTU~\cite{hstu}, RPG~\cite{rpg} and PROMISE~\cite{promise}.

\textbf{Evaluation Metrics.} 
We adopt two widely used evaluation metrics Recall@$K$ and NDCG@$K$ ($K = 5, 10$). 
Test set evaluation is conducted using the checkpoint with the best validation results.

\textbf{Implementation Details.}
To ensure a fair comparison, we use the same \texttt{sentence-t5-base}~\cite{sentence-t5} as the semantic encoder for items in both our experiments and reproductions, followed by quantization via RQ-VAE~\cite{rq-vae} or Res-KMeans~\cite{qarm}. We set the layer of SID $L=3$ and the codebook size $K = 256$. We employ the same T5-based encoder-decoder as our model architecture, with 4 layers and a dimension of 64. We fix the initial learning rate at 5e-4 with cosine decay. We train for up to 200 epochs with early stopping patience of 20. The beam size is consistently set to 20, and inference strategy selects pre-merge. All models are implemented in PyTorch.

Since offline public datasets only expose user-item interaction sequences without multi-behavior logs, we treat each ground-truth next item in the user sequence as a positive sample with implicit clicks, while negative samples are drawn from other items appearing within the same batch. 
This design ensures that a reward-guided mechanism with a single objective can still be maintained under public datasets.

\subsection{Industrial Scenario Experiment Settings}
\label{app:industry}

\textbf{Datasets.}
We train the model via continuous online learning based on logs from the Kuaishou live-streaming platform. Specifically, each day encompasses approximately 400 million users, 3 million creators, and billions of interaction records, including diverse behaviors such as clicks, long views, follows, and gifts. The data stream is organized as a real-time 30-second sliding window, and the detailed composition specifics can be referred to in~\cite{moment}.

\textbf{Baseline.}
We build upon OneLive~\cite{onelive}, a fully deployed generative recommendation service in our industrial environment, as our baseline. 
On this foundation, we incorporate various reward-aligned preference optimization methods for comparison, including RL approaches such as DPO~\cite{dpo} and GRPO~\cite{grpo}, as well as RGD guided by multiple objective reward combinations. The guidance signals encompass single-behavior corresponding scores: CTR, LVTR, WTR, GTR, and a multi-objective hybrid score LTR.

\textbf{Evaluation Metrics.}
We adopt HitRate and Mean Reciprocal Ranking (MRR) to qualify the generation ability.
Besides, we leverage Reward metric, following~\cite{onelive}, which employs a deployed ranking model to score and average the candidate items, thereby measuring the downstream pipeline's recognition of the generated outputs.

\subsection{Beam Size Expand}

\begin{table}[h]
  \caption{The comparison of results based on different generative model and beam size.}
  \label{tab:beam}
  \begin{tabular}{cccccc}
    \toprule
    \multirow{2.5}{*}{\textbf{Model}} & \multirow{2.5}{*}{\textbf{Beam Size}} & \multicolumn{2}{c}{\textbf{HitRate}} & \multicolumn{2}{c}{\textbf{CTR\_Reward}} \\
    \cmidrule(r){3-4} \cmidrule(r){5-6} & & @302 & @604 & @302 & @604 \\
    \midrule
    Base & 302 & 0.8021 & - & 0.0564 & -  \\
    Base & 604 & 0.8087 & 0.8188 & 0.0557 & 0.0550 \\
    RGD & 604 & 0.8146 & 0.8229 & 0.0570 & 0.0559 \\
    \bottomrule
  \end{tabular}
\end{table}

We further analyze the effect of beam size in Table~\ref{tab:beam}. 
For the base model, increasing the beam size brings only limited improvement in HitRate, while the CTR reward even decreases. 
In contrast, RGD achieves better HitRate under the same beam size and also improves CTR reward at @604. 
This indicates that reward guided decoding makes the enlarged beam more effective.
Instead of merely exploring more likely candidates, RGD can use the additional search space to preserve and select items with higher business value.

\end{document}